\documentclass[aps,prl, twocolumn, showpacs,superscriptaddress]{revtex4-1}
\usepackage[utf8]{inputenc}

\usepackage{amssymb,amsfonts,amsmath} 
\usepackage{graphicx,epsfig,psfrag}
\usepackage{color}
\usepackage{url}
\usepackage[breaklinks=true]{hyperref}
\usepackage{mathtools}
\usepackage{subfigure}
\hypersetup{
        colorlinks=true,
        citecolor    = blue
}
\usepackage{bookmark}
\usepackage{epic}
\usepackage{longtable}

\usepackage{bbm}
\usepackage{braket}
\usepackage{xcolor}
\usepackage{csquotes}
\usepackage{graphicx}
\usepackage{tikz}
\usepackage{extarrows}
\usetikzlibrary{trees}
\usetikzlibrary{decorations.pathmorphing}
\usetikzlibrary{decorations.markings}
\usetikzlibrary{positioning,arrows}
\usetikzlibrary{shapes.arrows}
\usepackage{leftidx}
\usepackage{framed}
\usepackage{tcolorbox}
\usepackage{bm}
\usepackage{MnSymbol}

\usepackage{romannum}
\AtBeginDocument{\pagenumbering{arabic}}

\tikzset{->-/.style={decoration={
  markings,
  mark=at position #1 with {\arrow{>}}},postaction={decorate}}} 

\usetikzlibrary{patterns}    
  
\usetikzlibrary{arrows,arrows.meta}
\newcommand\semiInt[1][1]{%
    \begin{tikzpicture}[scale=0.2]
        \coordinate (center) at (1.1,0.55);
        \draw[black,-{>[scale=0.6]}] (0, 0.0) + (center) arc (0:-180:0.5);
        \draw (0.1,0.55) --(1.1,0.55);
       $\int$
    \end{tikzpicture}
    }

\begin{document} 

\title{Fate of exceptional points in the presence of nonlinearities}

\author{Andisheh Khedri}
\affiliation{Institute  for  Theoretical  Physics,  ETH  Zurich,  8093  Zurich,  Switzerland}
\author{Dominic Horn}
\affiliation{Institute  for  Theoretical  Physics,  ETH  Zurich,  8093  Zurich,  Switzerland}
\author{Oded Zilberberg} 
\affiliation{Department of Physics, University of Konstanz, D-78457 Konstanz, Germany}
\begin{abstract}
The non-Hermitian dynamics of open systems deal with how intricate coherent effects of a closed system intertwine with the impact of coupling to an environment. The system--environment dynamics can then lead to so-called \textit{exceptional points}, which are the open-system marker of phase transitions, i.e., the closing of spectral gaps in the complex spectrum. Even in the ubiquitous example of the damped harmonic oscillator, the dissipative environment can lead to an exceptional point, separating between underdamped and overdamped dynamics at a point of critical damping. Here, we examine the fate of this exceptional point in the presence of strong correlations, i.e., for a nonlinear oscillator. By employing a functional renormalization group approach, we identify non-perturbative regimes of this model where the nonlinearity makes the system more robust against the influence of dissipation and can remove the exceptional point altogether. The melting of the exceptional point occurs above a critical nonlinearity threshold.
Interestingly, the exceptional point melts faster with increasing temperatures, showing a surprising flow to coherent dynamics when coupled to a warm environment. 
\end{abstract}

\maketitle

\textit{Introduction} -- Quantum mechanics and its Hamiltonian description focus on a closed system premise, exhibiting coherence that is the key resource for unlocking future quantum technologies~
\cite{Degen2017,Ladd2010}. In realistic situations, however, any quantum system is coupled to an environment that results in decoherence and dissipation~\cite{Breuer2002},  which generically poses an obstruction for quantum  applications. At the same time,  exploiting dissipation can result in a variety of advantageous novel effects, such as engineering of robust quantum communication~\cite{Duan2001}, dissipation-induced topological effects~\cite{Diehl2011}, and dissipative phase transitions~\cite{Zoller2010,Soriente2018,Soriente2021,Ferri2021}. 

To address such open quantum systems, Liouvillians are routinely used~\cite{Albert2016,Nori2019,Popkov2021}. Whereas closed systems have a real spectrum and a corresponding unitary time evolution, open systems are generally described by a complex spectrum~\cite{Rueter2010,El-Ganainy2018,Wang2019} and evolve with non-Hermitian operators. Similarly to phase transitions in closed systems, the open system exhibits critical points in parameter space, where the gap in the complex spectrum closes. Such critical points, which correspond to degeneracies in the open system, are known as  exceptional points (EPs)~\cite{Heiss2012}, with the most ubiquitous EP example manifesting for the damped harmonic oscillator (DHO) at critical damping~\cite{Fernandez2018}. In recent years, EPs have gained considerable attention due to the variety of the applications that they offer, e.g., in quantum sensing~\cite{Wiersig2014,Miri2019}, in photonics~\cite{Miri2019,Feng2017}, and in topology~\cite{BergholtzRevModPhys}. 

\begin{figure}[ht!]
    \centering
    \includegraphics[width=\columnwidth]{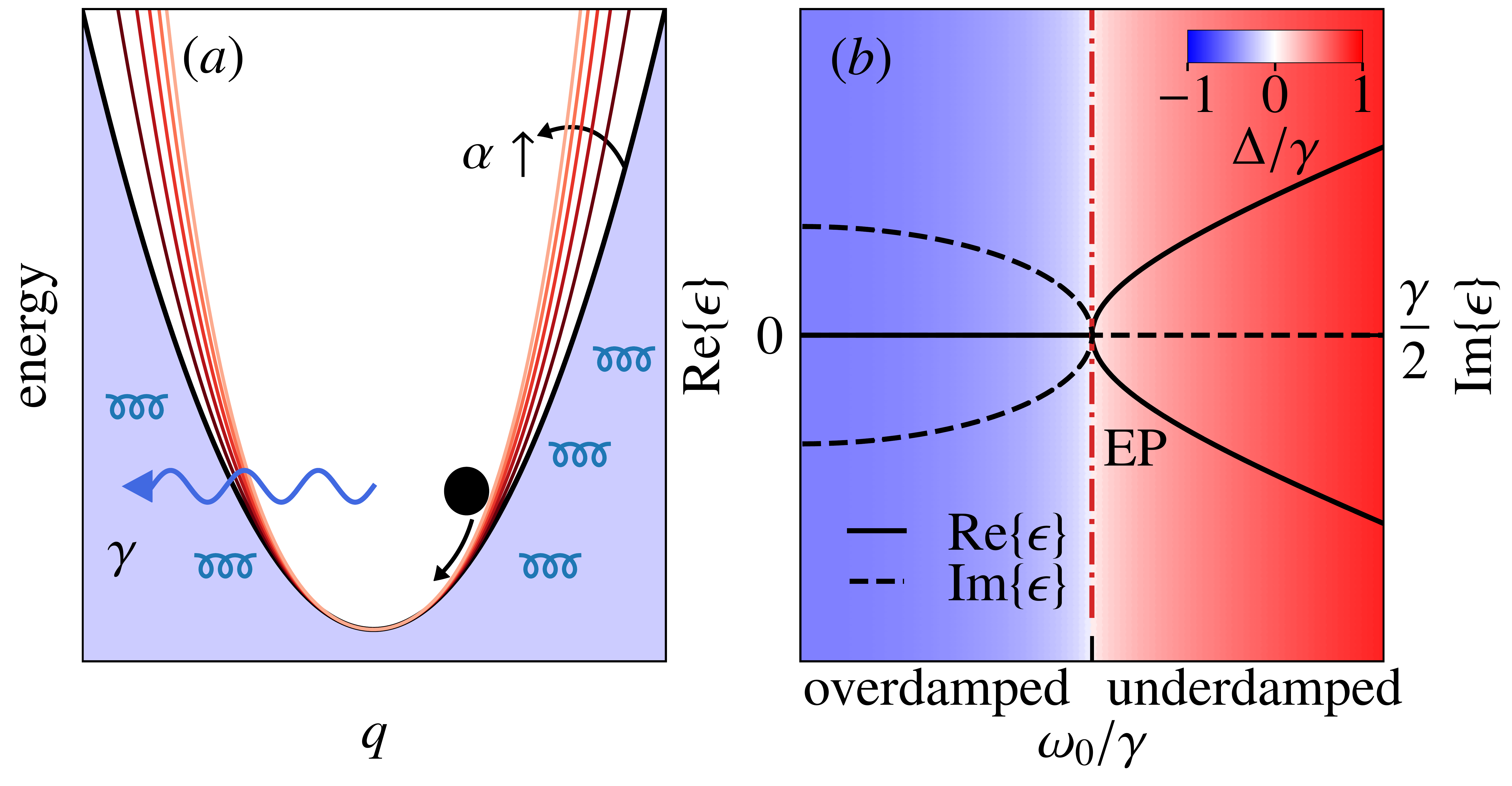}
    \caption{(a) Sketch of a massive particle moving around a minimum in a 1D potential-energy landscape with displacement $q$ [cf.~Eq.~\eqref{eq:EOM}]. The quadratic (black) potential is corrected by repulsive ($\alpha>0$) quartic nonlinearities (red lines depict different $\alpha$). The particle's motion is damped by the environment (blue bath with rate $\gamma$, wiggly blue arrow). (b) The real and imaginary parts of the characteristic exponent $\epsilon$ as a function of the oscillator's Q-factor, $\omega_0/\gamma$. The real part of $\epsilon$ in the underdamped region (red background) has two branches that merge together at an exceptional point, i.e., at critical damping $\gamma=2\omega_0$. In the overdamped case (blue background), the real part of $\epsilon$ vanishes. The background color depicts the order parameter $\Delta$.}
    \label{fig:fig1}
\end{figure} 

In both closed and open systems,
nonlinearities 
result in collective many-body effects that are oftentimes challenging to describe theoretically. They offer a plethora of rich strongly-correlated physics ranging from Kondo effects~\cite{hewson_1993,Stocker2022} to superfluidity~\cite{Khalatnikov1965,Vilchynskyy2013}, and to high-temperature superconductivity~\cite{Sharma2021}. One of the paradigmatic examples to see such many-body effects is the \textit{$\phi^4$ theory}~\cite{sachdev2011}; the building block of quantum field theory~\cite{Peskin1995} with implications across many fields, ranging from
superfluidity~\cite{Bogolyubov1947} to high-energy physics~\cite{Brandenberger}. To study a \textit{$\phi^4$ theory} in a closed system, renormalization group approaches are commonly employed~\cite{Wilson1983,Sinner2010} as conventional perturbative treatments  break down~\cite{Nozieres1964,Shi1998}.
In open systems, the competition between coherent processes and the incoherent forcing induced by the environment raises a fundamental question regarding the robustness of collective many-body effects against dissipation. Hence, the study of strong correlations in dissipative quantum systems is drawing much contemporary interest~\cite{Nakagawa2018,Louren2018,Ashida2017,Zhang2022,Grunwald2022}, albeit with a focus on how the nonlinearity acts as the source for the non-Hermitian system. 

In this work, we explore the fate of  exceptional points in nonlinear systems.  We introduce  nonlinearity as a quartic correction to the potential energy of a DHO, and find that the   \textit{$\phi^4$ theory}   washes away the dissipation-induced EP above a critical threshold. This implies that the impact of a dissipative environment can be suppressed with many-body interactions. Furthermore, we study the influence of thermal fluctuations in the system, and find that they amplify the impact of the nonlinearity. Thus, the EP can be removed by heating the system. We propose a scheme by which to measure our prediction in a plethora of systems, and expect that our result bears implications to a broad range of fields.

\textit{Model} -- 
We study a damped nonlinear oscillator (DNO) governed by the equation of motion
\begin{equation}
\ddot{q}+ \omega_0^2 q+ \gamma \dot{q} + \alpha q^3 = 0 \,,  
\label{eq:EOM}
\end{equation}
where $q$ is the displacement, dots mark differentiation with respect to time, $\omega_0$ is the bare frequency of the harmonic oscillator, $\gamma$ is the linear dissipation rate, and $\alpha$ is the Duffing nonlinearity. Equation~\eqref{eq:EOM} describes many physical systems, ranging from mechanical, electrical, and optical resonators~\cite{Landau_Lifshitz,Kleinert2004} to the inflation of the universe~\cite{Brandenberger}. As an example, we consider a massive particle oscillating in a quartic 1D potential energy while dissipating energy to the walls, see Fig.~\ref{fig:fig1}(a).

In the harmonic (DHO) limit ($\alpha=0$), we can employ the ansatz $q=e^{i\epsilon t}$ to obtain a quadratic characteristic equation for $\epsilon$, with solutions $\epsilon_{1,2}=i\gamma/2\pm\sqrt{\omega_0^2-(\gamma/2)^2}$, see Fig.~\ref{fig:fig1}(b). Whenever, $\epsilon_1\neq\epsilon_2$, i.e., $\gamma\neq 2\omega_0$, we thus obtain the corresponding basis for the harmonic motion $q_{1,2}=e^{i\epsilon_{1,2}t}$. We can  distinguish between two different phases: (i) an underdamped regime, $\gamma<2\omega_0$ ($\text{Re}\{\epsilon\}\neq 0$), with oscillatory behaviour that persists until the systems ``rings down'', and (ii) an overdamped regime,  $\gamma>2\omega_0$ ($\text{Re}\{\epsilon\}= 0$), where the system's energy is lost faster than the oscillation cycle. The two oscillation regimes meet at an EP when 
$\epsilon_1=\epsilon_2$ ($\gamma=2\omega_0$), i.e., where the ansatz does not yield a unique solution and the system is critically damped. In the following, we will use the order parameter $\Delta=\left[|\text{Re}\{\epsilon_1\}|-|\text{Im}\{\epsilon_1\}|+\gamma/2\right]/\gamma$ to continuously distinguish between the phases, i.e.,  in (i) $\Delta>0$, in (ii) $\Delta<0$, and at the EP $\Delta=0$.  

In the presence of the Duffing term, $\alpha\neq 0$, the nonlinear  Eq.~\eqref{eq:EOM} cannot be solved analytically for generic parameters~\cite{Kleinert2004}. Instead, various approximate methods are employed. For example, deep in the underdamped regime and for small $\alpha$, perturbative approaches such as the Poincar\'{e}-Lindstedt~\cite{Drazin1992} and Green's functions diagrammatics~\cite{Kleinert2004} were developed. 
In this work, we go beyond previous studies by considering the whole crossover from underdamped to overdamped regimes using a full quantum mechanical treatment. In particular, we address the impact of a small nonlinearity, $\alpha/ \text{max}\{\omega_0,\gamma\}^3\ll 1$, on the exceptional point appearing for the quantum DHO. To this end, we formulate the problem in the imaginary-time formalism~\cite{Altland2010}, \footnote{Note, that for the case of a dissipative environment where the fluctuation-dissipation relation holds, our treatment yields the same results as a Keldysh action analysis~\cite{kamenev2011field,Sieberer2016,supmat}.} where the action for the DNO takes the form
\begin{align}
S=\int_{0}^{\beta} d\tau~ \phi^*(\tau)
G_0^{-1}(\tau)
\phi(\tau)
+\alpha [\phi^*(\tau)\phi(\tau)]^2\,,
\label{eq:action_dho}
\end{align}
where $\phi^{(*)}$ is the quantum field operators corresponding to the position operator, and the free imaginary-time Green's function is defined as the solution to the equation of motion corresponding to the first three terms in Eq.~\eqref{eq:EOM}, namely, it satisfies $[-\frac{d^2}{d\tau^2}+\omega_0^2+i\gamma\frac{d}{d\tau}]G_0(\tau)=-\delta(\tau)$. 

\begin{figure}[t!]
    \centering
    \includegraphics[width=\columnwidth]{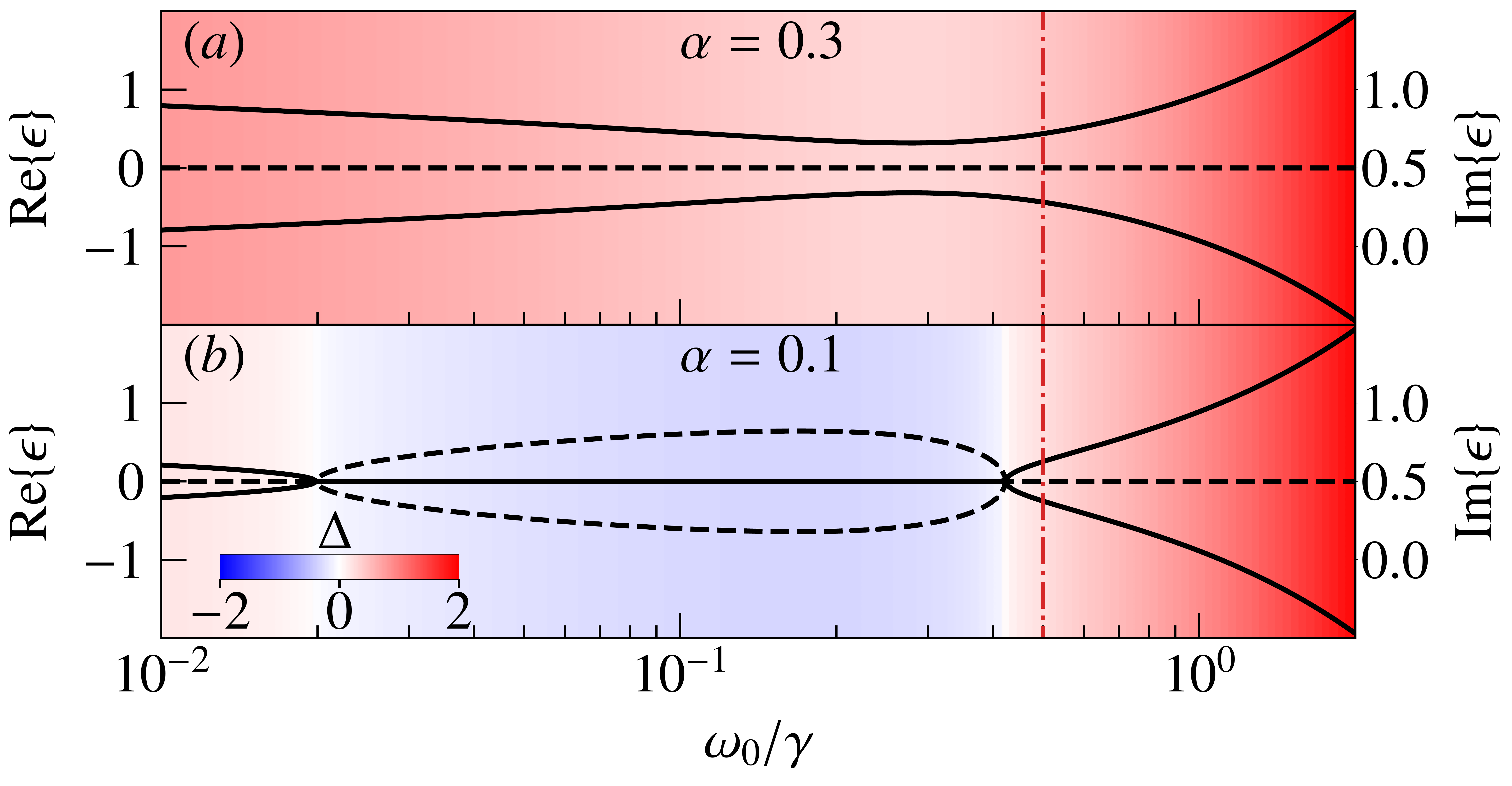}
    \caption{Renormalized excitation energy (solid lines) and life time (dashed lines), real and imaginary part of the poles of Green's function, respectively, using a one-loop perturbation [cf.~Eq.~\eqref{eq:self_energy_PT}] as a function of the $Q$-factor for a fixed dissipation strength $\gamma=1$. (a) $\alpha=0.3$, and (b) $\alpha=0.1$. The dashed-dotted vertical line marks the position of the EP in the DHO case, cf.~Fig.~\ref{fig:fig1}(b). The background color depicts the order parameter $\Delta$.}
    \label{fig:fig2}
\end{figure} 

In Matsubara space, $G(i\omega_n)=1/\beta\int_{0}^{\beta} d\tau ~ G(\tau)e^{i\omega_n\tau}$, where $\beta$ is the inverse temperature,  and the bare Green's function reads
\begin{equation}
G_0(i\omega_n)=-\frac{1}{\omega_n^2+\omega_0^2+\gamma|\omega_n|}\,,
\label{eq:free_G_dho}
\end{equation}
where $\omega_n=2\pi n/\beta$ denotes the bosonic Matsubara frequencies with $n\in \mathbb{N}$. Using analytic continuation, we can define the  retarded free Green's function $\mathcal{G}^{\rm R}_0(\omega)=G_0(i\omega_n\to \omega+i0^+)$, where the real and imaginary parts of its poles determine the energy and lifetime of excitations in the system, respectively. Analogous to our above discussion [cf.~Fig.~\ref{fig:fig1}], we identify the underdamped region, $\gamma<2\omega_0$, for which $\mathcal{G}^{\rm R}_0(\omega)$ contains poles that have both a real (frequency), $\rm{Re}\{\epsilon\}=\pm\sqrt{\omega_0^2-(\gamma/2)^2}$, and an imaginary  (lifetime) part, $\rm{Im}\{\epsilon\}=\gamma/2$. The underdamped region ends at the critically-damped point, $\gamma=2\omega_0$, for which the real part vanishes, i.e., at the exceptional point of this model.

\textit{Perturbation theory} -- 
To answer how the nonlinearity impacts the existence of exceptional points, we focus on the dressed imaginary-time Green's function,   $G(i\omega)\equiv\text{Tr}\{\phi(i\omega) \phi(-i\omega) \exp[-S]\}$ [cf.~Eq.~\eqref{eq:action_dho}]. Using a perturbation theory (PT) approach and employing Dyson's equation, we can write the dressed Green's function as $G(i\omega)=[[G_0(i\omega)]^{-1}-\Sigma]^{-1}$, where we defined the self-energy $\Sigma$ in terms of a perturbative expansion in the nonlinearity strength as~\cite{Bruus2004}  
\begin{equation}
\Sigma_{\rm PT}=\alpha \frac{1}{\beta}\sum_n [G_0(i\omega_n)]
e^{i\omega_n 0^+}+\mathcal{O}[\alpha^2]\,.
\label{eq:self_PT}
\end{equation}
Our treatment can be readily written in terms of  Feynman diagrams~\cite{supmat}, where we focus on the one loop correction (Hartree term). Plugging Eq.~\eqref{eq:free_G_dho} into Eq.~\eqref{eq:self_PT}, we first focus on the zero temperature limit, where the self-energy reads~\cite{supmat}
\begin{equation}
\Sigma_{\rm PT}=
\begin{cases}
\frac{\alpha}{2\sqrt{\omega_0^2-(\frac{\gamma}{2})^2}}\left[1-\frac{2}{\pi}\arctan\left(\frac{\gamma}{2\sqrt{\omega_0^2-(\gamma/2)^2}}\right)\right] & ;\,\frac{\gamma}{2}<\omega_0\,,\\
\frac{\alpha}{2\pi\sqrt{(\frac{\gamma}{2})^2-\omega_0^2}} \log\left|\frac{\gamma+2\sqrt{(\gamma/2)^2-\omega_0^2}}{\gamma-2\sqrt{(\gamma/2)^2-\omega_0^2}}\right| & ;\,\frac{\gamma}{2}>\omega_0\,,\\
\frac{2\alpha}{\pi\gamma} & ;\,\frac{\gamma}{2}=\omega_0\,,
\end{cases}
\label{eq:self_energy_PT}
\end{equation}
for the different regions of the model. It is important to note that in the overdamped regime, the perturbative self-energy diverges logarithmically, i.e.,  for $\omega_0\ll \gamma$, we have $\Sigma_{\rm PT}\approx \alpha/(\pi\gamma)\log[\gamma^2/\omega_0^2]$.

We obtain a dressed retarded Green's function  $\mathcal{G}^{\rm R}(\omega)=1/[\omega^2+i\gamma\omega-\omega_{\rm PT}^2]$ with the renormalized oscillator frequency defined as $\omega_{\rm PT}=\sqrt{\omega_0^2+\Sigma_{\rm PT}}$. Such a loop correction arises purely from quantum fluctuations of the oscillator. For $\alpha>0$, the renormalized frequency is always larger than the bare one, $\omega_{\rm PT}>\omega_0$ [cf.~Eq.~\eqref{eq:self_energy_PT}]. In other words, the nonlinearity increases the frequency of oscillations in the system.
Such a renormalization of the oscillation frequency can be understood intuitively: the positive nonlinearity increases the restoring force, i.e., it makes the potential energy steeper as a function of position, see Fig.~\ref{fig:fig1}(a). Crucially, for sufficiently large $\alpha$, we can have a situation where $\omega_0<\gamma/2$ but $\omega_{\rm PT}>\gamma/2$, such that the EP of the system is thus lifted, and the overdamped phase cannot be realized, see Fig.~\ref{fig:fig2}(a). For a small nonlinearity, instead, the EP manifests albeit at a smaller $Q$-factor as compared to the linear case, see Fig.~\ref{fig:fig2}(b). This implies a critical value of nonlinearity $\alpha_{\rm crit}$ beyond which the exceptional point vanishes.

The perturbative results predict a non-monotonic behaviour of the excitation energy $\text{Re}\{\epsilon\}$ as a function of $\omega_0$  [cf.~Eq.~\eqref{eq:self_energy_PT} and Fig.~\ref{fig:fig2}], which implies a fragility of EPs to nonlinearities. Our result however is not physical, e.g., by decreasing the bare oscillation frequency, the system can go from overdamped to underdamped dynamics, see Fig.~\ref{fig:fig2}(b). Such unphysical behaviour combined with the logarithic divergence in the perturbative expansion [cf.~Eq.~\eqref{eq:self_energy_PT}], raise questions regarding the reliability of the PT and its prediction concerning EPs. This motivates applying a non-perturbative approach to reveal the true impact of nonlinearities on EPs, which we perform in the following.

\textit{Functional renormalization group} -- 
To cure the above logarithmic divergency, we apply a functional renormalization group approach, in which a flow parameter $\Lambda \in [0,\infty]$ is introduced in the free Green's function $G^\Lambda_{0}$,
and high-frequency degrees of freedom (compared to $\Lambda$) are integrated out~\cite{Metzner2012,Kopietz2010,Grunwald2022}.
Following this procedure, we are able to approach the overdamped region with care, and obtain a hierarchy of differential equations (flow equations) for the one-particle irreducible vertex functions, e.g., for the self-energy $\Sigma^{\Lambda}$~\cite{supmat}. 
Whereas this procedure is formally exact, we need to truncate the hierarchy of flow equations in order to have a tangible solution. Here, we use the so-called \textit{first-order truncation scheme}, which is controlled for weak nonlinearities, such that only the self-energy flows as  
\begin{multline}
\partial_{\Lambda}\Sigma^{\Lambda} =\\ -\frac{\alpha}{\beta} \sum_{n} G^\Lambda(i\omega_n) \left(\partial_\Lambda [G_0^{\Lambda}(i\omega_n)]^{-1}\right)G^\Lambda(i\omega_n) e^{i\omega_n 0^+}\,, 
\label{eq:flow_self_general}
\end{multline}
with the scale-dependent Green's function $G^\Lambda=[[G^\Lambda_{0}]^{-1}-\Sigma^{\Lambda}]^{-1}$.

\begin{figure}[t!]
    \centering
    \includegraphics[width=1.0\linewidth]{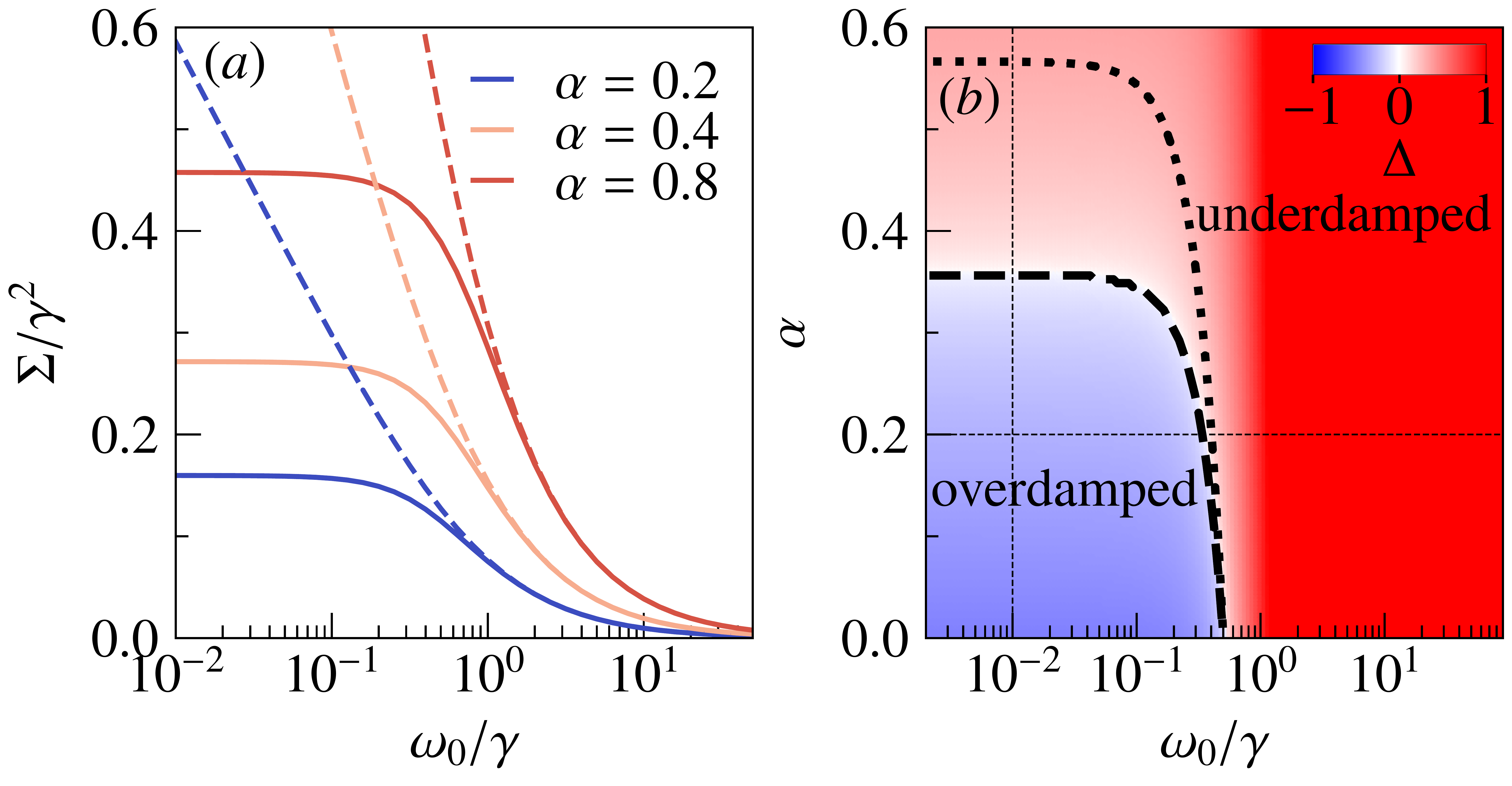}
    \caption{ (a) Comparison between the FRG [solid lines, cf.~Eq.~\eqref{eq:dno_flow}] and PT [dashed lines, cf.~Eq.~\eqref{eq:self_energy_PT}] self-energy as a function of the $Q$-factor with $\gamma=1$ for various nonlinearity strengths. (b) Phase diagram of the DNO, characterized by $\Delta$, as a function of nonlinearity $\alpha$ and $Q$-factor. The exceptional point (critical damping) manifests along the dashed line. The dotted line depicts the EP predicted by our analytical self-consistent approximation. Horizontal and vertical dotted lines mark the cuts plotted in Fig.~\ref{fig:TFRG}.
    }
    \label{fig:DHO_FRG}
\end{figure}

\begin{figure*}[t!]
    \centering
    \includegraphics[width=1.0\linewidth]{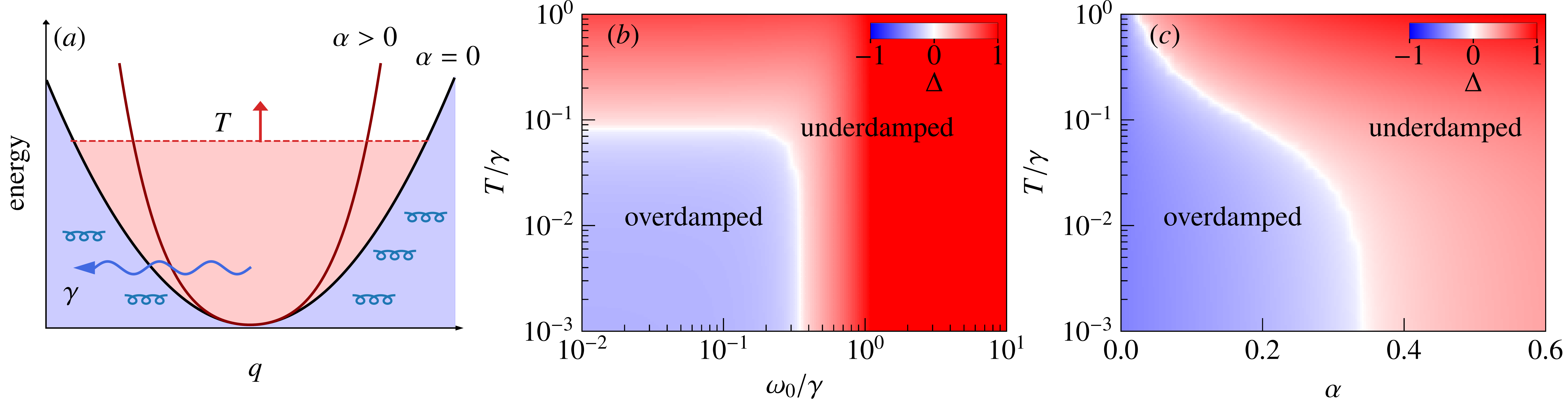}
    \caption{(a) Sketch of the potential landscape of a linear (black line) and nonlinear (red line) damped harmonic oscillator, cf.~Fig.~\ref{fig:fig1}(a). Increasing the temperature, expands the phase space that the trapped particle explores in the potential well (red fill and arrow). (b) The FRG resulting phase diagram [cf.~Eqs.~\eqref{eq:flow_self_general} and~\eqref{eq:T_dep_cuoff}] as a function of temperature and $Q$-factor with $\gamma=1$ and $\alpha=0.2$ [cf.~horizontal line in Fig.~\ref{fig:DHO_FRG}(b)]. (c) Same as (b), as a function of temperature and nonlinearity for $\gamma=1$ and $\omega_0=0.1$ [cf.~vertical line in Fig.~\ref{fig:DHO_FRG}(b)].}
    \label{fig:TFRG}
\end{figure*}

To compare with the perturbative result in Eq.~\eqref{eq:self_energy_PT}, we first focus on the zero temperature limit of the flow~\eqref{eq:flow_self_general}. To this end, we consider the so-called \textit{sharp cutoff scheme}, where the free Green's function is scaled as $G_{0}^\Lambda(i\omega)=\Theta(|\omega|-\Lambda)G_0(i\omega_n)$~\cite{Metzner2012}. Plugging  $G_{0}^\Lambda(i\omega)$ into Eq.~\eqref{eq:flow_self_general} simplifies the flow to
\begin{equation}
\partial_{\Lambda}\Sigma^{\Lambda}=-\frac{\alpha}{\pi}\frac{1}{\omega_0^2+\Lambda^2+\gamma\Lambda+\Sigma^{\Lambda}}\,.  
\label{eq:dno_flow}
\end{equation}
We solve this flow numerically and denote the solution $\Sigma_{\rm FRG}$. In Fig.~\ref{fig:DHO_FRG}(a), we plot $\Sigma_{\rm FRG}$ and compare the result with the perturbative treatment [cf.~Eq.~\eqref{eq:self_energy_PT}]. Crucially, we observe that the logarithmic divergence  is regularized in the FRG approach. This is due to the presence of the self-energy in the right-hand side of Eq.~\eqref{eq:dno_flow} that saturates the self-energy to a finite value for $\omega_0\ll \gamma$. We proceed and insert the FRG self-energy into the retarded Green's function, and obtain a corresponding regularized-renormalization of the oscillator frequency $\omega_{\rm FRG}\equiv\sqrt{\omega_0^2+\Sigma_{\rm FRG}}$. 

Similarly to the results presented in Fig.~\ref{fig:fig2}, we systematically study the fate of the EP as a function of $\alpha$ and $\omega_0/\gamma$ and collect the outcome into a phase diagram, see Fig.~\ref{fig:DHO_FRG}(b). Importantly, the unphysical phase predicted by the PT at $\omega_0\ll\gamma$  vanishes [cf.~Fig.~\ref{fig:fig2}(b)]. 
We can now draw a conclusion concerning the fate of the critical-damping EP. At weak nonlinearity, the exceptional point is pushed to smaller  $\omega_0/\gamma$, and the overdamped region is shrinking. This fits with our intuition that the effective oscillation frequency grow with increasing $\alpha$, and counters the dissipation. Interestingly, beyond a critical nonlinearity value $\alpha_{\rm crit}$, the EP is fully removed from the system. Physically, beyond the critical value,  dissipation cannot counter the coherent forcing enacted by the quartic potential. This is the main result of our paper. 

To gain more analytical insight, we can estimate the regularized self-energy in the overdamped case by solving the self-consistent equation $\Sigma_{\rm SC}=\alpha/(\pi\gamma)\log[\gamma^2/(\omega_0^2+\Sigma_{\rm SC})]$~\cite{supmat}. Fixing $\Sigma_{\rm SC}+\omega_0^2\equiv(\gamma/2)^2$ for a given $\omega_0$ and $\gamma$, we then find the nonlinearity for which the EP is realized as $\alpha_{\rm SC}/\gamma^3=\pi[(1/4)-(\omega_0/\gamma)^2]/\log 4$. The estimate agrees well with the exact result for critical values $\omega_0\sim\gamma$ and is independent of the bare frequency in the limit of $\omega_0\ll\gamma$, in agreement with the FRG result, see Fig.~\ref{fig:DHO_FRG}(b).

\textit{Finite Temperature} -- 
Our treatment thus far dealt with the fate of the EP at zero temperature, $T=0$. With increasing temperature, thermal fluctuations excite the DNO, i.e., they allow the trapped particle in the potential well to explore positions further away from the bottom of the well, see Fig.~\ref{fig:TFRG}(a). Fortunately, our Matsubara formalism in Eqs.~\eqref{eq:self_PT} and~\eqref{eq:flow_self_general} allows for the exploration of the  DNO's EP physics at finite temperatures. 
From Eq.~\eqref{eq:self_PT}, we observe that also at $T\neq 0$, the PT yields in unphysical divergences~\cite{supmat}, which we want to resolve with the aid of the FRG approach. 

To set up FRG for $T>0$, we modify the sharp cutoff with a smeared one~\cite{Enss2005}, i.e., the scale-dependent Green's function now reads $G_0^\Lambda(i\omega_n)=G_{0}(i\omega_n)\Theta_T(|\omega_n|-\Lambda)$ with \textit{smeared cutoff} 
\begin{equation}
\Theta_T\big(|\omega_n|-\Lambda\big)=
\begin{cases}
0 &|\omega_n|-\Lambda\le-\pi T\\
\frac{1}{2}+\frac{|\omega_n|-\Lambda}{2\pi T}&\Big||\omega_n|-\Lambda\Big| < \pi T\\
1 & |\omega_n|-\Lambda\ge \pi T\\
\end{cases}.
\label{eq:T_dep_cuoff}
\end{equation}
With this cutoff inserted into Eq.~\eqref{eq:flow_self_general}, we numerically solve the resulting truncated flow equation~\cite{supmat} and obtain a phase diagrams akin to that of Fig.~\ref{fig:DHO_FRG}(b), see Figs.~\ref{fig:TFRG}(b) and (c). We observe that with increasing temperatures, the overdamped regions shrink, and specifically that $\alpha_{\rm crit}$ goes down with temperature $T$. This is a result of an interesting interplay between thermal fluctuations and the impact of the nonlinearity on the system. Specifically, we can understand this interplay using the following semi-classical picture: thermal fluctuations kick the particle away from the basin where the quadratic (harmonic) approximation holds, i.e., it explores large displacements where nonlinearities are more pronounced. As the nonlinearity $\alpha>0$ increases the renormalized oscillator frequency and thus reduces the overdamped region, we observe the melting of the overdamped region with temperature, see Fig.~\ref{fig:TFRG}(b). This results in an exotic case where rising temperatures lead to more coherent motion in the system by enhancing the impact of the nonlinearity. 

Exceptional points mark observable criticality in open systems. We predict an exotic interplay between such environment-induced phenomenon and the intrinsic many-body interactions in the closed system. A possible route to measuring our predictions involves starting in an overdamped system, $Q < 1/2$, and tuning the nonlinearity of the device or by increasing the temperature, see Figs.~\ref{fig:DHO_FRG} and~\ref{fig:TFRG}. As the overdamped system is time-translation invariant and the resulting state has a characteristic frequency, the EP-melting in a macroscopic mode purports a so-called continuous time crystal~\cite{heugel2019classical,heugel2022role,sacha_time_2020,kessler2020continuous}. Due to the ubiquitous nature of our model, our prediction can be realized in a broad range of physical systems, and even pertain to open questions in cosmology. Our methodology opens the path for future studies of exceptional point physics via higher-order renormalization group studies, as well as generalizations to other settings where such physics manifests.

\begin{acknowledgments}
\textit{Acknowledgments} -- We thank A. Eichler, M. Soriente, T. L. Heugel, J. del Pino, T. Donner, and T. Giamarchi for illuminating discussions. We acknowledge financial support from the Swiss National Science Foundation (SNSF) through project 190078, and from the Deutsche Forschungsgemeinschaft (DFG) - project number 449653034.
\end{acknowledgments}
\bibliography{ref}
\clearpage

\setcounter{equation}{0}

\setcounter{figure}{0}

\setcounter{enumi}{0}
\renewcommand{\theequation}{S\arabic{equation}}

\begin{widetext}
\section*{Supplementary material}
\subsection{Action of a damped oscillator}
For completeness, we provide here the derivation of the imaginary-time action of a damped harmonic oscillator (DHO)~\cite{Kleinert2004}. We begin with the action of a harmonic oscillator (HO) whose position satisfies the equation of motion $\ddot{q}+ \omega_0^2 q=0$, or equivalently whose Hamiltonian reads  $H_0=\frac{\dot{q}^2}{2}+\omega_0^2 q^2$. The partition function of the HO is  $\mathcal{Z}=\text{Tr}\{e^{-\beta H_0}\}$, which can be formulated in the path-integral formalism as $\mathcal{Z}=\int \mathcal{D}x~ e^{-S_{0}}$, with the imaginary-time action
\begin{equation}
S_0=\int_{0}^{\beta}d\tau~\left[\frac{\dot{q}(\tau)^2}{2}+\omega_0^2 q^2(\tau)\right]   
=\int_0^{\beta}d\tau~ q(\tau) g^{-1}_0(\tau,\tau^\prime)~ q(\tau)\,,
\end{equation}
where the oscillator Green's function $g_0$ satisfies
\begin{equation}
\left[-\partial_\tau^2+\omega_0^2\right]g_0(\tau,\tau^\prime)
=-\delta(\tau-\tau^\prime)\,.
\end{equation}
We now investigate how such a Green's function gets modified if the HO is in contact with a heat bath. The latter has a dispersion $\Omega_k$, composed of infinitely-many oscillators with the action
\begin{equation}
S_{\rm bath}=\sum_{k} \int_{0}^{\infty} d\tau~x_k(\tau)~g^{-1}_{{\rm b},k}(\tau-\tau^\prime)~ x_k(\tau)\,, 
\end{equation}
where $g_{\rm b}$ is the bath's Green's function, satisfying $\left[-\partial_\tau^2+\omega_k^2\right]g_{{\rm b},k}(\tau,\tau^\prime)=-\delta(\tau-\tau^\prime)$. In Matsubara frequency space, it is defined as $g_{{\rm b},k}(i\omega_n)=\int d\tau e^{i\omega_n(\tau-\tau^\prime)} g_{{\rm b},k}(\tau-\tau^\prime)$, and takes the form
\begin{equation}
g_{{\rm b},k}(i\omega_n)=\frac{1}{(i\omega_n)^2-\Omega_k^2}  \,.  
\label{eq:bath_green}
\end{equation}
The action corresponding to the coupling of the HO to the bath reads
\begin{equation}
S_{\rm coup}=\sum_{k}\lambda_k \int_{0}^{\beta}d\tau~ x_k(\tau) ~ q(\tau)\,,
\end{equation}
with $\lambda_k$ being the coupling constant to the $k-th$ oscillator in the bath. 

We can integrate out the bath's degrees of freedom and obtain
\begin{equation}
\int \mathcal{D} q~\int \mathcal{D}[x_k]~ e^{-\left[S_0+S_{\rm bath}+S_{\rm coup}\right]}=\int \mathcal{D} q~ e^{-\int_{0}^{\beta}\int_{0}^{\beta} d\tau^\prime d\tau~q(\tau)~G_0^{-1}(\tau,\tau^\prime)~q(\tau^\prime)}\,,
\end{equation}
with
\begin{equation}
G_0(\tau,\tau^\prime)=\frac{1}{g^{-1}_0(\tau,\tau^\prime)-\Sigma_{\rm bath}(\tau,\tau^\prime)} \,, 
\end{equation}
and the bath-induced self-energy
\begin{equation}
\Sigma_{\rm bath}(\tau,\tau^\prime)=\sum_k |\lambda_k|^2 g_{b}(\tau,\tau^\prime)
=\frac{1}{\beta} \sum_n ~ \Sigma_{\rm bath}(i\omega_n)~e^{-i\omega_n(\tau-\tau^\prime)}\,.
\label{eq:effective_cor}
\end{equation}
Thus, we can identify 
\begin{equation}
\Sigma_{\rm bath}(i\omega_n)=\sum_k \frac{|\lambda_k|^2}{\omega_n^2+\Omega_k^2}\,,
\label{eq:bath_slef}
\end{equation}
by plugging Eq.~\eqref{eq:bath_green} in Eq.~\eqref{eq:effective_cor}. Defining the bath spectral density as~\cite{Kleinert2004}
\begin{equation}
\rho_{\rm b}(\omega) = 2\pi \sum_k \frac{|\lambda_k|^2}{2\Omega_k} \delta(\omega-\Omega_k)\,,
\end{equation}
we can rewrite the bath-induced self-energy as 
\begin{equation}
\Sigma_{\rm bath}(i\omega_n)=
\int_{0}^{\infty} \frac{d\omega}{2\pi} ~\rho_{\rm b}(\omega) ~\frac{2\omega}{\omega_n^2+\omega^2}
=\int_0^{\infty}d\omega~\frac{\rho_{\rm b}(\omega)}{\omega}\Big[1-\frac{\omega_n^2}{\omega_n^2+\omega^2}\Big]\,.  
\label{eq:bath_slef_int}
\end{equation}
Furthermore, considering an Ohmic bath with bandwidth $\omega_c$, the bath spectral density reads
\begin{equation}
\rho_{\rm b}^{\rm Ohmic}(\omega)=\gamma\frac{2\omega}{1+(\omega/\omega_c)^2}\,,  
\label{eq:Ohmic_spec}
\end{equation}
and Eq.~\eqref{eq:bath_slef_int} becomes
\begin{equation}
\Sigma_{\rm Ohmic}(i\omega_n)= 
-\frac{\gamma\omega_c\omega_m}{2}\sum_{s,s^\prime=\pm}\int_{-\infty}^{\infty}\frac{d\omega}{2\pi}\frac{s s^\prime}{(\omega+s i\omega_c)(\omega+s^\prime i\omega_n)}
=\gamma|\omega_n|\frac{\omega_c}{|\omega_n|+\omega_c}\,,
\end{equation}
corresponding to Ohmic dissipation.

\subsection{Action of a damped harmonic oscillator in Keldysh formalism}

Commonly, an open system scenario is better treated using a Keldysh action formalism~\cite{kamenev2011field}. We present here, for completeness, that in our case of a single oscillator coupled to a single Ohmic bath, the Matsubara description suffices~\cite{kamenev2011field}. The bath-induced self-energy in the keldysh formalism reads
\begin{equation}
\Sigma_{\rm bath}^{\rm R/A/K}(\omega)=\sum_{k} |\lambda_k|^2 g^{\rm R/A/K}_{\rm b} (\omega) \,,   
\label{eq:bath_induced_self_RAK}
\end{equation}
where the bath's Green's function are
\begin{align}
g_0^{\rm R/A}(\omega)=\frac{1}{(\omega\pm i\eta)^2-\Omega_k^2}\,,\quad\quad\mathrm{and}\quad\quad
g_0^{\rm K}(\omega)=[\delta(\omega-\Omega_k)-\delta(\omega+\Omega_k)]\text{coth}(\beta\omega/2)\,.    
\label{eq:bath_green_keldysh}
\end{align}
We can identify that the fluctuation-dissipation theorem holds for the bath, i.e., 
\begin{equation}
g_0^{\rm K}(\omega)=[g^{\rm R}(\omega)-g^{\rm A}(\omega)][1+n_{b}(\omega)]\,,    
\end{equation}
where $n_{b}(\omega)=1/[e^{\beta\omega}-1]$ is the bosonic distribution function, and $\beta$ the inverse temperature.
Plugging Eq.~\eqref{eq:bath_green_keldysh} into Eq.~\eqref{eq:bath_induced_self_RAK}, the Keldysh component of the bath-induced self-energy becomes
\begin{equation}
\Sigma^{\rm K}_{\rm bath}=-i\, \text{coth}(\beta\omega/2)\,\rho_{\rm b}(\omega)\,,    
\end{equation}
with the definition
\begin{equation}
\rho_{\rm b}(\omega)=2\pi\sum_{k}\frac{|\lambda_k|}{2\Omega_k}
\left[ \delta(\omega-\Omega_k)-\delta(\omega+\Omega_k) \right]\,,    
\end{equation}
and the retartded/advanced bath-induced self-energy reads
\begin{equation}
\Sigma^{\rm R/A}_{\rm bath}(\omega)=\int_{-\infty}^{\infty} \frac{d\omega^\prime}{2\pi}  ~\rho_{b}(\omega)~\frac{\omega^\prime}{(\omega\pm i \eta)-{\omega^\prime}^2}\,.  
\end{equation}
If we consider an Ohmic spectral density of the form Eq.~\eqref{eq:Ohmic_spec}, we obtain
\begin{equation}
\Sigma^{\rm R/A}_{\rm Ohmic}=-i\gamma\omega~\frac{\omega_c}{\omega_c+\omega}\,.    
\end{equation}
Note that in the limit of $\omega_c\to \infty$, we recover the fluctuation-dissipation relation for the bath-induced self-energy
\begin{equation}
\Sigma_{\rm Ohmic}^{\rm K}(\omega)=[\Sigma_{\rm Ohmic}^{\rm R}(\omega)-\Sigma_{\rm Ohmic}^{\rm A}(\omega)] [1+2n_{b}(\omega)]\,.   
\end{equation}
\subsection{Perturbation theory for the Damped Nonlinear Oscillator}

In this section, we outline the calculation of the self-energy for the damped nonlinear oscillator (DNO) within first-order perturbation theory, i.e., one loop correction (Hartree term), as it is shown diagramatically in Fig.~\ref{fig:diagrams}(a). In written form, it reads
\begin{equation}
\Sigma_{\rm PT}(1;1^\prime)=\sum_{2,2^\prime}G_0(2;2^\prime)L_2(1,2^\prime;1^\prime,2)\,,    
\label{eq:PT_general}
\end{equation}
where $G_0$ is the free Green's function, $L_2$ is the two-particle vertex function, and $n=1,1^\prime,2,2^\prime$ enumerates the degrees of freedom. For the DNO, (i) the  $G_0(n;n^\prime)$ in (Matsubara) frequency space depends on a single frequency, and (ii) the vertex function is frequency independent. Due to (ii), we can replace  $L_2$ by the nonlinearity $\alpha$, and obtain Eq.~(4) in the main text. 
\begin{figure}[h!]
\begin{tikzpicture}[scale=0.5]

\tikzset{
    partial ellipse/.style args={#1:#2:#3}{
        insert path={+ (#1:#3) arc (#1:#2:#3)}
    }
}

\draw(-10.0,4.0)node{(a)};

\draw[line width=0.4mm] (-8,2) circle (1cm);
\draw[thick,->-=0.5,line width=0.4mm](-10.0,2.0)--(-9.0,2.0);
\draw[thick,->-=0.5,line width=0.4mm](-7.0,2.0)--(-6.0,2.0);
\draw(-8.0,2.0)node{$\Sigma$};

\draw (-5.0,2.0)node{$=$};

\draw[thick,->-=0.5,line width=0.4mm](-4.0,2.0)--(-2.0,2.0);
\draw[thick,->-=0.5,line width=0.4mm](-2.0,2.0)--(0.0,2.0);
\draw(-2,1.0)node{$L_2$};
\draw[thick,->-=0.7,line width=0.4mm] (-2,3.05) [partial ellipse=270:-90:1.0cm and 1.0cm];
\fill[white] (-2.0,2.0) ellipse (0.3cm and 0.3cm);
\fill[draw=black,pattern=north west lines] (-2.0,2.0) ellipse (0.3cm and 0.3cm);

\draw(4.0,4.0)node{(b)};

\fill[black!100!white] (6.0,3.5) ellipse (0.15cm and 0.15cm);
\draw[line width=0.4mm] (6,2) circle (1cm);
\draw[thick,->-=0.5,line width=0.4mm](4.0,2.0)--(5.0,2.0);
\draw[thick,->-=0.5,line width=0.4mm](7.0,2.0)--(8.0,2.0);
\draw(6.0,2.0)node{$\Sigma^\Lambda$};

\draw (9.0,2.0)node{$=$};

\draw[line width=0.4mm] (12,2) circle (1cm);
\draw[thick,->-=0.5,line width=0.4mm](10.0,2.0)--(11.0,2.0);
\draw[thick,->-=0.5,line width=0.4mm](13.0,2.0)--(14.0,2.0);
\draw(12.0,2.0)node{$L_2^\Lambda$};
\draw[thick,->-=0.7,line width=0.4mm] (12,2.9) [partial ellipse=210:-30:1.0cm and 1.0cm];
\draw[line width=0.4mm](11.8,3.6)--(12.2,4.2);
\end{tikzpicture}
\caption{(a) The diagramatic expansion of the self-energy up to second order in nonlinearity $\alpha$ within perturbation theory, cf.~Eq.~(4) in the main text. The solid lines represent the bare propagator cf.~Eq.~(3) in the main text, and the hashed circle represent the bare vertex. (b) Diagramatic representation of the flow equation of the self-energy within the FRG approach. On the left-hand side, the dot marks the derivative of the self-energy with respect to the flow parameter $\Lambda$. On the right-hand side, the crossed-out line denotes the single-scale
propagator $S^\Lambda$. In the first order truncation scheme, we replace the scale-dependant vertex function with the bare one, i.e., $L_2^\Lambda=L_2$, see Eq.~(6) in the main text.}
\label{fig:diagrams}
\end{figure}
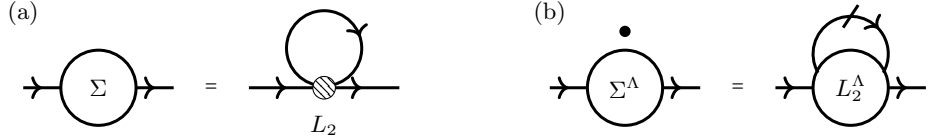
Plugging the bare Green's function [Eq.~(3) in the main text] into the perturabtive expansion [Eq.~(4) in the main text]  results in    
\begin{equation}
\Sigma_{\rm PT}=\frac{\alpha}{\beta}\sum_n e^{i\omega_n\tau} \prod_{s=\pm} \frac{1}{i\omega_n+i(\gamma/2)\text{sgn}(\omega_n)+s\Omega_\gamma}\,,    
\label{eq:self_dho_c}
\end{equation}
with $\omega_n=2n\pi/\beta$, $n\in\mathbb{N}$, $\tau>0$ a convergence factor~\cite{Bruus2004}, and $\Omega_\gamma\equiv\sqrt{\omega_0^2-(\gamma/2)^2}$, which is real in the underdamped case and purely imaginary in the overdamped region. 

We emphasize that Eq.~\eqref{eq:self_dho_c} is valid for arbitrary temperatures $T$, and all the information about the temperature dependence of the self-energy is encoded in the Matsubara summation over $n$. In order to evaluate such a summation, we rewrite the Matsubara sums in terms of contour integrals~\cite{Bruus2004}. Specifically, we can show
\begin{align}
\semiInt \frac{dz}{2\pi i}\frac{n_b(z) e^{z\tau}}{z-i\gamma/2\pm \Omega_\gamma}
=\int_{+\infty}^{-\infty} \frac{d\omega}{2\pi i} \frac{e^{\omega\tau} n_{b}(\omega)}{\omega-i\gamma/2\pm\Omega_\gamma}\,,
\label{eq:contour}
\end{align}
as the integral over the semi-circle with radius $R\to\infty$ vanishes because $\tau\in(0,\beta)$, and hence $n_b(z)e^{\tau z}\to 0$ for $\text{Re}(z)\to -\infty$.
Furthermore, using the Residue theorem, we can rewrite
\begin{align}
\semiInt \frac{dz}{2\pi i}\frac{n_b(z) e^{z\tau}}{z-i\gamma/2\pm \Omega_\gamma}
= \frac{1}{\beta} \sum_{n<0} \frac{e^{i\omega_n\tau}}{i\omega_n-i\frac{\gamma}{2}\pm \Omega_\gamma}\,,
\label{eq:contour_res}
\end{align}
since $\text{Res}[n_b(z)]|_{z=i\omega_n}=1/\beta$.
Complex conjugating the integrand in Eqs.~\eqref{eq:contour} and~\eqref{eq:contour_res}, and considering the half-circle in the upper complex plane, we can show analogous equalities, which we summarize as
\begin{align}
\frac{1}{\beta}\sum_{n\gtrless 0} \frac{e^{i\omega_n\tau}}{i\omega_n+ i\frac{\gamma}{2}\pm\Omega_\gamma}=\pm \int_{-\infty}^{\infty} \frac{d\omega}{2\pi i}\frac{n_{b}(\omega)e^{\omega\tau}}{\omega+ i\frac{\gamma}{2}\pm\Omega_\gamma}\,.
\label{eq:kpo_c_int}
\end{align}

Therefore the perturbative self-energy of Eq.~\eqref{eq:self_dho_c} becomes
\begin{equation}
\Sigma_{\rm PT}=\frac{\alpha\gamma}{2\Omega_\gamma}\int_{-\infty}^{\infty}\frac{d\omega}{2\pi} 
e^{\omega\tau} n_{b}(\omega)\sum_{s=\pm}\frac{s}{(\omega+s\Omega_\gamma)^2+\gamma^2/4}\,.
\label{eq:PT_self_integral}
\end{equation}

\subsubsection{Zero temperature limit}

In the limit of zero temperature, we can approximate the bosonic distribution function as $\lim_{\beta\to\infty}n_{b}(\omega)=-\theta(-\omega)$, and therefore the integral in Eq.~\eqref{eq:PT_self_integral} can be evaluated as
\begin{equation}
\Sigma^{T\to 0}_{\rm PT}=\frac{\alpha\gamma}{2\Omega_\gamma}\lim_{\tau\to 0}\int_{-\infty}^{0}\frac{d\omega}{2\pi} 
e^{\omega\tau}\sum_{s=\pm}\frac{-s}{(\omega+s\Omega_\gamma)^2+\gamma^2/4}\,,
\label{eq:PT_self_integral_zeroT}
\end{equation}
which simplifies to Eq.~(5) in the main text for different regions of the DNO.

\subsubsection{Finite temperatures}
At finite temperatures, we can evaluate the integral in Eq.~\eqref{eq:PT_self_integral} numerically, see outcome in Fig.~\ref{fig:PT_finite_t}. The renormalized oscillator frequency is independent of $T$ at $T\ll\omega_0$, i.e., when the thermal energy is not sufficient to excite the oscillator. However, at $T\gg\omega_0$, we see a linear dependence on temperature. At such high temperatures, we can approximate the bosonic distribution function in Eq.~\eqref{eq:PT_self_integral} as $\lim_{\beta\to 0}n_{b}(\omega)=1/(\beta\omega)$, which explains the linear dependence of the self-energy on temperature, see inset to Fig.~\ref{fig:PT_finite_t}. Note that the perturbative self-energy diverges at $\omega_0\to T$, as shown in Fig.~\ref{fig:FRG_finite_t}(b).

\begin{figure*}[t!]
    \centering
    \includegraphics[width=1.0\linewidth]{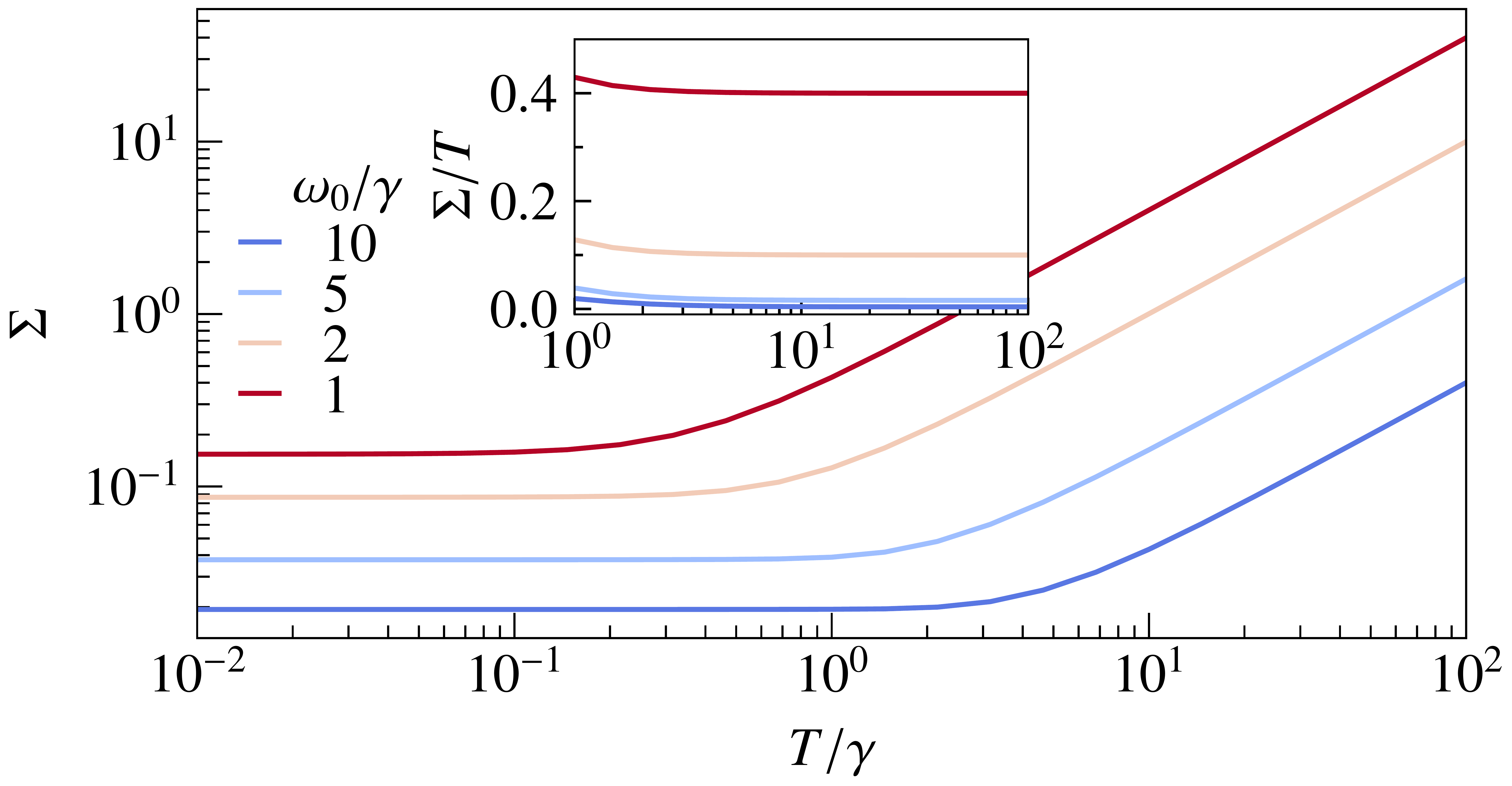}
    \caption{Perturbative self-energy [cf.~Eq.~\eqref{eq:PT_self_integral}] as a function of temperature for various oscillator frequencies. The inset shows that the self-energy depends linearly on temperature for $T\gg\omega_0$.
    }
    \label{fig:PT_finite_t}
\end{figure*}

\subsection{Functional renormalization group analysis for the DNO}
\label{sec:TFRG}

In this section, we elaborate on the derivation of the functional renormalization group (FRG) flow equations for the DNO and their numerical solutions presented in the main text. In the FRG approach~\cite{Metzner2012}, a scale parameter $\Lambda\in[0,\infty]$ is introduced. The corresponding scale-dependant free propagator $G_0^\Lambda$ is defined such that it vanishes in the beginning of the flow $G_0^{\Lambda\to\infty}=0$, and it approaches the bare Green's function at the end of the flow $G_0^{\Lambda\to 0}=G_0$.   
Generally, the flow equation for the self-energy reads [see Fig.~\ref{fig:diagrams}(b)]
\begin{equation}
\partial_{\Lambda}\Sigma^\Lambda(1;1^\prime)=-\sum_{2;2^\prime} S^{\Lambda}(2;2^\prime)L_2^\Lambda(1,2^\prime;1^\prime,2)\,,   
\label{eq:flow_self_general_s}
\end{equation}
where $S^\Lambda$ is the single-scale propagator, defined as
\begin{equation}
S^\Lambda(1;1^\prime)=\sum_{2,2^\prime}G^\Lambda(1;2) \left(\partial_\Lambda [G_0^{\Lambda}(2;2^\prime)]^{-1}\right)G^\Lambda(2^\prime;1^\prime)\,,    
\label{eq:single_scale_prop}
\end{equation}
with $G^\Lambda=[(G^\Lambda_0)^{-1}-\Sigma^\Lambda]^{-1}$, and $L_2^\Lambda$ is a scale-dependent two-particle vertex function. The latter generally has its own flow equation, see Ref.~\cite{Metzner2012}. We focus here, instead, on the first-order truncation scheme, where the two-particle vertex function stays constant and frequency-independent during the flow, i.e., $L_2^\Lambda\equiv L_2$. Hence, Eq.~\eqref{eq:flow_self_general_s} simplifies to Eq.~(6) in the main text, which dictates the flow of the self-energy at any temperature, provided that the scale-dependant free propagator $G_0^\Lambda$ is chosen appropriately. 

\subsubsection{The zero temperature limit}
At $T=0$, the distance between Matsubara frequencies vanishes and we can treat them as a continuous variable $\omega_n\to\omega$. In this limit, we work with the sharp cut-off scheme, where $G_0^\Lambda=\theta(|\omega|-\Lambda)G_0$, and thereby Eq.~\eqref{eq:single_scale_prop} becomes
\begin{equation}
S^\Lambda(i\omega)=\frac{\delta(|\omega|-\Lambda) G^{-1}_0(\omega)}{[G^{-1}_0(\omega)-\theta(|\omega|-\Lambda)\Sigma^{\Lambda}]^2}\,.   
\end{equation}

\begin{figure*}[t!]
    \centering
    \includegraphics[width=1.0\linewidth]{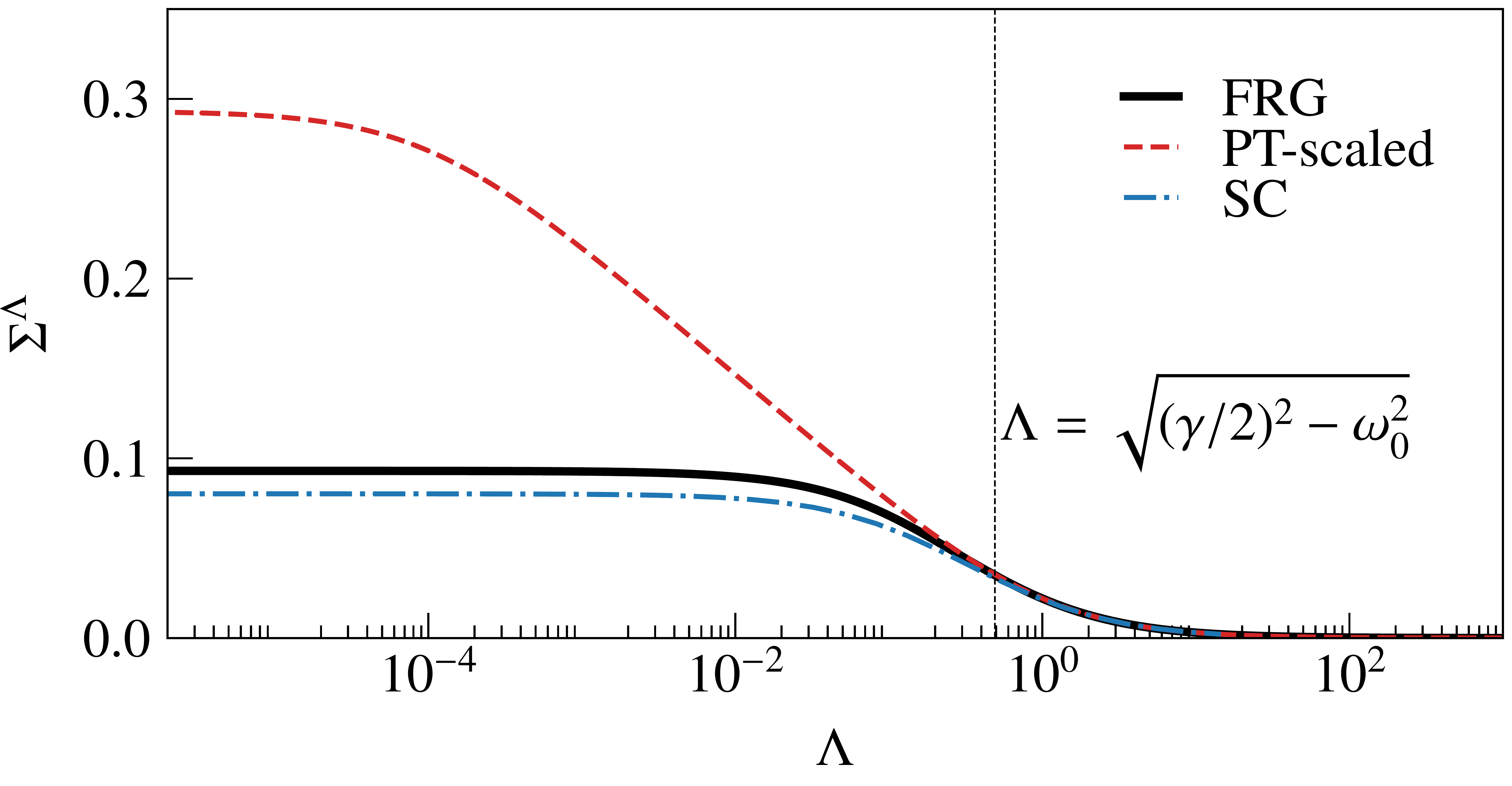}
    \caption{The flow of the self-energy at zero temperature as a function of the scale parameter $\Lambda$. We compare FRG [cf.~Eq.~\eqref{eq:dno_flow_s}, solid line] with the perturbative solution [cf.~Eq.~\eqref{eq:flow_per}, dashed line], marked as "PT-scaled", and with the self-consistent approximate solution [cf.~Eq.~\eqref{eq:Flow_self}, dashed-dotted], marked as "SC", for $\gamma=1$, and $\alpha=0.1$. We are in the overdamped situation $\omega_0/\gamma=0.01$. The dashed vertical line marks the scale corresponding to $-i\Omega_\gamma=\sqrt{(\gamma/2)^2-\omega_0^2}$.
    }
    \label{fig:Flow}
\end{figure*}

Furthermore, employing the Morris Lemma~\cite{Morris94}, we obtain
\begin{equation}
S^\Lambda(i\omega)=\frac{\delta(|\omega|-\Lambda)}{G_0(i\omega)}\int_{0}^{1}dx~ \frac{1}{[G_0^{-1}(i\omega)-x\Sigma^\Lambda]^2}=\frac{\delta(|\omega|-\Lambda)}{G_0^{-1}(i\omega)-\Sigma^\Lambda}\,,   
\end{equation}
which simplifies  Eq.~(6) to Eq.~(7) in the main text, i.e., to
\begin{equation}
\partial_{\Lambda}\Sigma^{\Lambda}=-\frac{\alpha}{\pi}\frac{1}{\omega_0^2+\Lambda^2+\gamma\Lambda+\Sigma^{\Lambda}}\,.  
\label{eq:dno_flow_s}
\end{equation}

It is important to note that if we neglect the self-energy on the right-hand side of the flow~\eqref{eq:dno_flow_s}, and perform the integration from $\Lambda=\infty$ to $\Lambda$, we obtain
\begin{equation}
\Sigma^\Lambda_{\rm PT}= \frac{-\alpha}{2\pi\sqrt{(\gamma/2)^2-\omega_0^2}}
\log\left[\frac{\Lambda+(\gamma/2)-\sqrt{(\gamma/2)^2-\omega_0^2}}{\Lambda+(\gamma/2)+\sqrt{(\gamma/2)^2-\omega_0^2}}\right]\,.  
\label{eq:flow_per}
\end{equation}
By setting $\Lambda=0$, Eq.~\eqref{eq:flow_per} boils down to the perturbative result~\eqref{eq:self_dho_c}, when the Matsubara sum can be replaced by an integral in the zero temperature limit.

For the overdamped case $\gamma\gg\omega_0$,  Eq.~\eqref{eq:flow_per} further simplifies to
\begin{equation}
\Sigma_{\rm PT}^\Lambda=\frac{\alpha}{\pi\gamma}\log\left[\frac{\Lambda+\gamma}{\Lambda+\omega_0^2/\gamma}\right]\,,
\label{eq:flow_per_overdamped}
\end{equation}
which for $\Lambda=0$ results in $\Sigma_{\rm PT}=\alpha/(\pi\gamma)\log[\gamma^2/\omega_0^2]$, as discussed in the main text. The numerical solution to the flow~\eqref{eq:dno_flow_s} is shown in Fig.~\ref{fig:Flow} for the overdamped case. At large scales, i.e., $\Lambda>\sqrt{(\gamma/2)^2-\omega_0^2}$, the self-energy is small compared to all other scales, such that the perturbative solution~\eqref{eq:flow_per} is a good approximation.

\begin{figure*}[t!]
    \centering
    \includegraphics[width=1.0\linewidth]{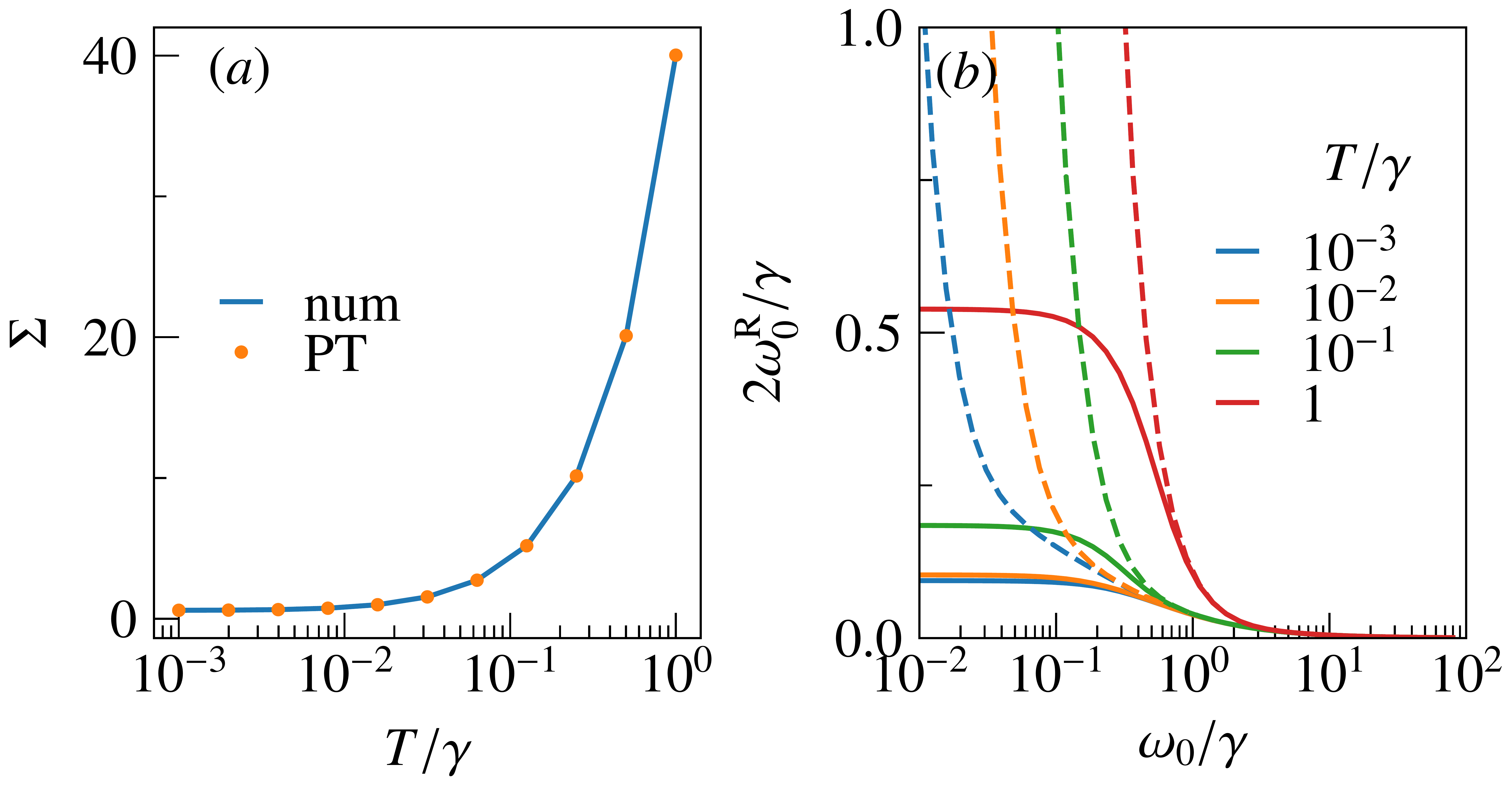}
    \caption{(a) Comparison of the perturbative self-energy obtained by numerically integrating the integral in Eq.~\eqref{eq:PT_self_integral}, marked as "PT", with the numerical solution to Eq.~\eqref{eq:flow_T} without the inclusion of the self-energy on the right-hand side for $\omega_0/\gamma=0.1$, and $\alpha=0.1$. The latter should be equivalent to the perturbative results.
    (b) The renormalized oscillator frequency as a function of the bare frequency $\omega_0/\gamma$ obtained from perturbation theory (dashed lines), and the FRG approach (solid lines) at various temperatures for $\gamma=1$ and $\alpha=0.1$. 
    }
    \label{fig:FRG_finite_t}
\end{figure*}

To gain more analytical insight to the behaviour of the flow at small scales, we can replace $\omega_0^2$ in Eq.~\eqref{eq:flow_per_overdamped} by its renormalized value, i.e., $\omega_0^2+\Sigma$. We thus obtain a self-consistent equation that can be solved as
\begin{equation}
\Sigma^\Lambda_{\rm SC}=-\Lambda\gamma-\omega_0^2+\frac{\alpha}{\pi\gamma}\text{ProductLog}\left[
(\pi\gamma^2/\alpha)(\gamma+\Lambda)\exp\{\gamma\pi(\omega_0^2+\Lambda\gamma)/\alpha\}
\right]\,,
\label{eq:Flow_self}
\end{equation}
where $\text{ProductLog}(z)$ is the solution for $w$ in $z=w\exp(w)$. The comparison between this approximate solution and the FRG results in shown in Fig.~\ref{fig:Flow}. At the end of the flow, i.e., $\Lambda=0$, the self-consistent equation results in
\begin{equation}
\Sigma_{\rm SC}=\frac{\alpha}{\pi\gamma}\log\left[\frac{\gamma^2}{\omega_0^2+\Sigma_{\rm SC}}\right]\,.
\end{equation}
Note that such a self-consistent equation regularizes the logarithmic divergence appearing in perturbation theory. This approximation is valid in the overdamped case, and for small nonlinearity such that $\Sigma\ll\gamma^2$. 

For a given $\omega_0$, and $\gamma$, we can find the nonlinearity $\alpha_{\rm SC}$ for which the EP is realized, by fixing $\Sigma_{\rm SC}+\omega_0^2=\gamma^2/4$. Furthermore, by setting $\Lambda=0$ in Eq.~\eqref{eq:Flow_self}, we obtain
\begin{equation}
\text{ProductLog}\left[(\pi\gamma^3/\alpha_{\rm SC})\exp\{\pi\gamma\omega_0^2/\alpha_{\rm SC}\}\right]=\frac{\pi\gamma^3}{4\alpha_{\rm SC}}\,,
\end{equation}
which implies
\begin{equation}
\alpha_{\rm SC}=\pi\gamma^3 \frac{\frac{1}{4}-\frac{\omega_0^2}{\gamma^2}}{\log(4)}\,.
\end{equation}
This self-consistent approximation predicts that for $\omega_0\ll\gamma$ the critical nonlinearity, beyond which the EP cannot be realized is independent of $\omega_0$, in agreement with the FRG result, as discussed in the main text.

\subsubsection{Finite temperatures}
At finite temperatures, following the procedure introduced in Ref.~\cite{Enss2005}, we introduce the scale-dependent Green's function $G_0^\Lambda(i\omega_n)=G^{0}(i\omega_n)\Theta_T(|\omega_n|-\Lambda)$, with the smeared cut-off 
defined in Eq.~(8) in the main text.
Therefore, the single-scale propagator $S^\Lambda=G^\Lambda (\partial_\Lambda [G^\Lambda_0]^{-1})G^\Lambda$ becomes 
\begin{equation}
S^\Lambda(i\omega_n)=
\begin{cases}
\frac{1}{2\pi T}\frac{G_0^{-1}(i\omega_n)}{G_0^{-1}(i\omega_n)-(\frac{1}{2}-\frac{|\omega_n|-\Lambda}{2\pi T})\Sigma^\Lambda}&||\omega_n|-\Lambda|\le\pi T\\
0 & \text{elsewhere}\,.
\end{cases}
\end{equation}
As the Matsubara frequencies have spacing of $2\pi/\beta$, at a given scale $\Lambda$, there is only one $m\in \mathbb{Z}$, for which $\omega_m$ lies in the interval $[\Lambda-\pi T,\Lambda+ \pi T]$. This removes the summation in  Eq.~\eqref{eq:flow_self_general_s}, and the flow equation for self-energy becomes
\begin{equation}
\partial_\Lambda \Sigma^{\Lambda}=\frac{-\alpha}{2\pi T}\sum_{s=\pm} 
\frac{\omega_m^2+\omega_0^2+\gamma|s\omega_m|}{[\omega_m^2+\omega_0^2+\gamma|s\omega_m|-(\frac{1}{2}-\frac{|s\omega_m|-\Lambda}{2\pi T})\Sigma^\Lambda]^2}\,,
\label{eq:flow_T}
\end{equation}
which can be integrated with standard numerical routines. Similar to the zero temperature limit, we can re-obtain the perturbative results by neglecting the self-energy on the right-hand side of Eq.~\eqref{eq:flow_T}, as is shown in Fig.~\ref{fig:FRG_finite_t}(a). The inclusion of the self-energy, however, will regularize the logarithmic divergencies appearing in the perturbative expansion, see Fig.~\ref{fig:FRG_finite_t}(b). 

\end{widetext}
\end{document}